\documentclass[aps,twocolumn,showkeys,showpacs,preprintnumbers,prd,superscriptaddress,nofootinbib,10pt]{revtex4-1}
\usepackage{graphicx,epsf,bm,amsmath,amsfonts,amssymb,epstopdf,natbib,hyperref,color,appendix,verbatim,multirow,bm,diagbox}
\hypersetup{colorlinks=true,urlcolor=blue,citecolor=blue,linkcolor=blue,menucolor=blue,anchorcolor=blue,filecolor=blue}

\newcolumntype{?}{!{\vrule width 3pt}}

\date{\today}

\makeatletter
\newcommand{\fmarki}{\ensuremath{\alpha}}
\newcommand{\fmarkii}{\ensuremath{\beta}}
\newcommand{\fmarkiii}{\ensuremath{\gamma}}
\newcommand{\fmarkiv}{\ensuremath{\delta}}
\newcommand{\fmarkv}{\ensuremath{\epsilon}}
                
\def\@fnsymbol#1{{\ifcase#1\or \fmarki\or \fmarkii\or \fmarkiii\or \fmarkiv\or \fmarkv\or \else\@ctrerr\fi}}
\makeatother

\begin{document}

\title{Multidimensionality of the Hubble tension: the roles of $\Omega_m$ and $\omega_c$}

\author{Davide Pedrotti}
\email{davide.pedrotti-1@unitn.it}
\affiliation{Department of Physics, University of Trento, Via Sommarive 14, 38123 Povo (TN), Italy}
\affiliation{Trento Institute for Fundamental Physics and Applications (TIFPA)-INFN, Via Sommarive 14, 38123 Povo (TN), Italy}

\author{Jun-Qian Jiang}
\email{jiangjunqian21@mails.ucas.ac.cn\\junqian.jiang@studenti.unitn.it}
\affiliation{School of Physical Sciences, University of Chinese Academy of Sciences, No.19(A) Yuquan Road, Beijing 100049, China}
\affiliation{Department of Physics, University of Trento, Via Sommarive 14, 38123 Povo (TN), Italy}

\author{Luis A. Escamilla}
\email{l.a.escamilla@sheffield.ac.uk}
\affiliation{School of Mathematics and Statistics, University of Sheffield, Hounsfield Road, Sheffield S3 7RH, United Kingdom \looseness=-1}

\author{Simony Santos da Costa}
\email{simony.santosdacosta@unitn.it}
\affiliation{Department of Physics, University of Trento, Via Sommarive 14, 38123 Povo (TN), Italy}
\affiliation{Trento Institute for Fundamental Physics and Applications (TIFPA)-INFN, Via Sommarive 14, 38123 Povo (TN), Italy}

\author{Sunny Vagnozzi}
\email{sunny.vagnozzi@unitn.it}
\affiliation{Department of Physics, University of Trento, Via Sommarive 14, 38123 Povo (TN), Italy}
\affiliation{Trento Institute for Fundamental Physics and Applications (TIFPA)-INFN, Via Sommarive 14, 38123 Povo (TN), Italy}

\begin{abstract}
\noindent The Hubble tension is inherently multidimensional, and bears important implications for parameters beyond $H_0$. We discuss the key role of the matter density parameter $\Omega_m$ and the physical cold dark matter density $\omega_c$. We argue that once $\Omega_m$ and the physical baryon density $\omega_b$ are calibrated, through Baryon Acoustic Oscillations (BAO) and/or Type Ia Supernovae (SNeIa) for $\Omega_m$, and via Big Bang Nucleosynthesis for $\omega_b$, any model raising $H_0$ requires raising $\omega_c$ and, under minimal assumptions, also the clustering parameter $S_8$. We explicitly verify that this behaviour holds when analyzing recent BAO and SNeIa data. We argue that a calibration of $\Omega_m$ as reliable and model-independent as possible should be a priority in the Hubble tension discussion, and an interesting possibility in this sense could be represented by galaxy cluster gas mass fraction measurements.
\end{abstract}

\maketitle

\section{Introduction}
\label{sec:introduction}

The ability of the simple, six-parameter $\Lambda$CDM cosmological model to account for a wide range of cosmological observations remains the main reason for its continued success. Yet, with the increase in precision of cosmological observations over the past decade, this success has been challenged by various tensions whose significance keeps growing: an example is the mismatch between the value of the Hubble constant $H_0=(67.36 \pm 0.54)\,{\text{km}}/{\text{s}}/{\text{Mpc}}$ inferred from Cosmic Microwave Background (CMB) data within $\Lambda$CDM~\cite{Planck:2018vyg}, and a number of direct measurements of the same quantity, including but not limited to the SH0ES Cepheid-calibrated Type Ia Supernovae (SNeIa) local distance ladder measurement of $H_0=(73.04 \pm 1.04)\,{\text{km}}/{\text{s}}/{\text{Mpc}}$~\cite{Riess:2021jrx}, see e.g.\ Refs.~\cite{Verde:2019ivm,DiValentino:2020zio,DiValentino:2021izs,Perivolaropoulos:2021jda,Schoneberg:2021qvd,Shah:2021onj,Abdalla:2022yfr,DiValentino:2022fjm,Hu:2023jqc,Verde:2023lmm} for reviews. Explanations based on systematics (e.g.\ Refs.~\cite{Efstathiou:2020wxn,Mortsell:2021nzg,Mortsell:2021tcx,Wojtak:2022bct,Freedman:2024eph}) are growing ever more challenging to defend, which is why serious consideration has been given to the possibility that the Hubble tension calls for new physics beyond $\Lambda$CDM (with no claims as to completeness, see Refs.~\cite{Anchordoqui:2015lqa,Karwal:2016vyq,Zhao:2017urm,Benetti:2017juy,Mortsell:2018mfj,Vagnozzi:2018jhn,Kumar:2018yhh,Yang:2018euj,Banihashemi:2018oxo,Guo:2018ans,Graef:2018fzu,Banihashemi:2018has,Agrawal:2019lmo,Li:2019yem,Yang:2019nhz,Vagnozzi:2019ezj,Visinelli:2019qqu,DiValentino:2019ffd,Escudero:2019gvw,DiValentino:2019jae,Niedermann:2019olb,Sakstein:2019fmf,Ye:2020btb,Hogg:2020rdp,Ballesteros:2020sik,Alestas:2020mvb,Jedamzik:2020krr,Ballardini:2020iws,DiValentino:2020evt,Banerjee:2020xcn,Niedermann:2020dwg,Gonzalez:2020fdy,Braglia:2020auw,RoyChoudhury:2020dmd,Brinckmann:2020bcn,Alestas:2020zol,Gao:2021xnk,Alestas:2021xes,Karwal:2021vpk,Cyr-Racine:2021oal,Akarsu:2021fol,Niedermann:2021ijp,Saridakis:2021xqy,Sen:2021wld,Herold:2021ksg,Odintsov:2022eqm,Heisenberg:2022lob,Heisenberg:2022gqk,Sharma:2022ifr,Ren:2022aeo,Nunes:2022bhn,Nojiri:2022ski,Schoneberg:2022grr,Joseph:2022jsf,Gomez-Valent:2022bku,Moshafi:2022mva,Odintsov:2022umu,Banerjee:2022ynv,Alvarez:2022wef,Ge:2022qws,Akarsu:2022typ,Gangopadhyay:2022bsh,Schiavone:2022wvq,Gao:2022ahg,Brinckmann:2022ajr,Khodadi:2023ezj,Dahmani:2023bsb,Ben-Dayan:2023rgt,deCruzPerez:2023wzd,Ballardini:2023mzm,Yao:2023ybs,Gangopadhyay:2023nli,Zhai:2023yny,SolaPeracaula:2023swx,Gomez-Valent:2023hov,Ruchika:2023ugh,Adil:2023exv,Frion:2023xwq,Akarsu:2023mfb,Escamilla:2023oce,Petronikolou:2023cwu,Sharma:2023kzr,Ben-Dayan:2023htq,Ramadan:2023ivw,Fu:2023tfo,Efstathiou:2023fbn,Montani:2023ywn,Lazkoz:2023oqc,Forconi:2023hsj,Sebastianutti:2023dbt,Benisty:2024lmj,Stahl:2024stz,Shah:2024rme,Giare:2024smz,Montani:2024pou,Co:2024oek,Akarsu:2024eoo,Yadav:2024duq,Nozari:2024wir,Dwivedi:2024okk,Montani:2024xys,Escamilla:2024xmz,Perivolaropoulos:2024yxv,Yao:2024kex} for examples).

Nevertheless, focusing on $H_0$ can only reveal part of the story and is, at best, misleading. A key role in the Hubble tension is in fact played by Baryon Acoustic Oscillation (BAO) measurements: once the sound horizon at baryon drag $r_d$ is calibrated to the $\Lambda$CDM value -- either from the CMB or through the Big Bang Nucleosynthesis (BBN) determination of the physical baryon density $\omega_b$ -- BAO can be combined with uncalibrated SNeIa measurements to build an \textit{inverse distance ladder}. This yields a value of $H_0$ compatible with the ``low'' \textit{Planck} $\Lambda$CDM one, yet in principle independent of any CMB data. Another way of seeing this is to note that the distance-redshift diagrams of SH0ES-calibrated SNeIa, and BAO calibrated through the $\Lambda$CDM value of $r_d$, are mutually completely inconsistent despite probing the same $0.1 \lesssim z \lesssim 2$ redshift range (see e.g.\ Fig.~1 of Ref.~\cite{Poulin:2024ken}, Fig.~1 of Ref.~\cite{Bousis:2024rnb}, as well as Refs.~\cite{Raveri:2019mxg,Benevento:2020fev,Camarena:2021jlr,Camarena:2023rsd,Raveri:2023zmr}). Given that BAO are sensitive to the combination $H_0r_d$, it becomes clear that solving the Hubble tension while not altering the SNeIa absolute magnitude $M_B$ necessarily requires a decrease in $r_d$. This, in turn, calls for new physics operating before recombination~\cite{Bernal:2016gxb,Addison:2017fdm,Lemos:2018smw,Aylor:2018drw,Schoneberg:2019wmt,Knox:2019rjx,Arendse:2019hev,Efstathiou:2021ocp,Cai:2021weh,Keeley:2022ojz}.~\footnote{See Refs.~\cite{Akarsu:2022lhx,Tutusaus:2023cms,Escamilla:2024ahl} for possible caveats regarding this conclusion.}

However, even looking at $H_0$ and $r_d$ is still not the end of the story, as models that aim to solve the Hubble tension are inevitably subject to several other constraints. Lowering $r_d$, therefore, is just a part of what a successful model should do, and various other cosmological quantities such as the age of the Universe $t_U$~\cite{Jimenez:2019onw,Bernal:2021yli,Vagnozzi:2021tjv,Wei:2022plg,Capozziello:2023ewq}, the physical matter density $\omega_m$~\cite{Pogosian:2020ded,Jedamzik:2020zmd,Bernal:2021yli}, and the fractional matter density parameter $\Omega_m$~\cite{Lin:2019htv,Lin:2021sfs,Bernal:2021yli,Sakr:2023hrl,Baryakhtar:2024rky} appear to play a key role in the Hubble tension discourse. It has therefore become increasingly clear that the Hubble tension is inherently a multi-dimensional problem, as recently emphasized by Ref.~\cite{Poulin:2024ken}, where the denomination ``cosmic calibration tension'' was suggested.

Our work fits in this context by further exploring the role played by two important cosmological parameters: the fractional matter density parameter $\Omega_m$, and the physical dark matter (DM) density parameter $\omega_c=\Omega_ch^2$, with $h$ the dimensionless Hubble constant. Assuming both $\Omega_m$ and the physical baryon density $\omega_b$ are somehow calibrated, one can easily show that an increase in $H_0$ \textit{must} be accompanied by an increase in $\omega_c$. Under minimal assumptions concerning the primordial power spectrum, this inevitably leads to a potentially problematic increase in the clustering parameter $S_8$~\cite{Poulin:2024ken}. It so happens that $\Omega_m$ and $\omega_b$ can in fact be calibrated in a fairly robust way, through a combination of BAO and uncalibrated SNeIa in the former case, and BBN in the latter. In the rest of our work we discuss these points in more detail, explicitly demonstrating on real data how an increase in $H_0$ is accompanied by an increase in $\omega_c$ (and $S_8$), while studying the impact of the $\Omega_m$ calibration and arguing that a high-fidelity, as model-independent as possible calibration of $\Omega_m$, should be a major priority in the quest towards solving the Hubble tension.

The rest of this paper is then organized as follows. We begin by discussing in more detail the role of $\Omega_m$ and $\omega_c$ in the context of the Hubble tension in Sec.~\ref{sec:multidimensionality}. In Sec.~\ref{sec:data} we discuss our datasets and methodology. We present our results in Sec.~\ref{sec:results}, and critically discuss them in Sec.~\ref{sec:discussion}. Finally, in Sec.~\ref{sec:conclusions} we draw concluding remarks.

\section{Multidimensionality of the Hubble tension}
\label{sec:multidimensionality}

The important role of $\Omega_m$ and $\omega_c$ in the Hubble tension can most easily be understood by expressing the fractional matter density parameter as follows:
\begin{eqnarray}
\Omega_m = \frac{\Omega_bh^2+\Omega_ch^2+\Omega_\nu h^2}{h^2} = \frac{\omega_b+\omega_c+\omega_\nu}{h^2}\,.
\label{eq:omegam}
\end{eqnarray}
where $\Omega_m$, $\Omega_b$, $\Omega_c$, and $\Omega_{\nu}$ are the matter, baryon, cold DM, and (massive) neutrino (fractional) density parameters, $h \equiv H_0/(100\,{\text{km}}/{\text{s}}/{\text{Mpc}})$ is the reduced Hubble parameter, and $\omega_m$, $\omega_b$, $\omega_c$, and $\omega_{\nu}$ are the matter, baryon, cold DM, and (massive) neutrino physical density parameters respectively. Note that in writing Eq.~(\ref{eq:omegam}) we are explicitly assuming a spatially flat Universe. In what follows, we will treat the sum of the neutrino masses $\sum m_{\nu}$ as known and fixed to $0.06\,{\text{eV}}$, so that $\omega_{\nu}=\sum m_{\nu}/94.13\,{\text{eV}} \lesssim 0.0015$ is also known and fixed.~\footnote{Given the currently very tight upper limits on $\sum m_{\nu}$~\cite{Vagnozzi:2017ovm,Wang:2024hen,Naredo-Tuero:2024sgf,Du:2024pai,Jiang:2024viw}, $\omega_{\nu}$ will not play a major role in what follows.} Let us now assume that both $\Omega_m$ and $\omega_b$ can be calibrated (by BAO+SNeIa and BBN respectively, as we shall discuss shortly), and can therefore be considered approximately constant within uncertainties. From Eq.~(\ref{eq:omegam}) we then see that an increase in $h$ is necessarily accompanied by an increase in $\omega_c$:
\begin{equation}
\omega_c = \Omega_m h^2 - (\omega_b + \omega_\nu) \,.
\label{eq:omegac}
\end{equation}
For small variations $\delta\omega_c$ and $\delta h$, we can therefore expecting the following to hold:
\begin{eqnarray}
\frac{\delta\omega_c}{\omega_m} \approx 2 \frac{\delta h}{h}\,,
\label{eq:deltac}
\end{eqnarray}
which can be rewritten as follows:
\begin{eqnarray}
\frac{\delta \omega_c}{\omega_c} = \frac{\delta\omega_c}{\omega_m}\frac{\omega_m}{\omega_c} \approx 2 \left ( 1+\frac{\omega_b+\omega_\nu}{\omega_c} \right ) \frac{\delta h}{h}\,.
\label{eq:deltaomegacdeltah}
\end{eqnarray}
For reference, taking the best-fit cosmological parameters inferred from a fit to the \textit{Planck} 2018 TTTEEE+lowE+lensing likelihoods~\cite{Planck:2018vyg}, for which $\omega_b=0.022$, $\omega_c=0.120$, and $\omega_m=0.143$, we find that Eq.~(\ref{eq:deltaomegacdeltah}) reduces to the following:
\begin{eqnarray}
\frac{\delta \omega_c}{\omega_c} \approx 2.38\frac{\delta h}{h}\,,
\label{eq:deltacplanck2018}
\end{eqnarray}
which shows that the fractional increase in the physical DM density $\omega_c$ must be larger than twice the fractional increase in the Hubble constant $H_0$.

We remark once more that the earlier arguments hinge upon calibrations for $\Omega_m$ and $\omega_b$. We now discuss in more detail how this is achieved. To set the stage, in what follows we work within a spatially flat Friedmann-Lema\^{i}tre-Robertson-Walker (FLRW) metric, and assume that the Etherington distance-duality relation (DDR)~\cite{Etherington:1933ghw} holds, in agreement with recent data~\cite{Santos-da-Costa:2015kmv,Hogg:2020ktc,Renzi:2021xii,Liu:2023hdz,Qi:2024acx}. Within our assumptions, the transverse comoving distance $D_M$ is given by the following:
\begin{eqnarray}
D_M(z) = \int_0^z\, \frac{dz'}{H(z')} = \frac{1}{H_0}\int_0^z\, \frac{dz'}{E(z')}\,,
\label{eq:dm}
\end{eqnarray}
where we denote by $E(z) \equiv H(z)/H_0$ the unnormalized expansion rate. Transverse, line-of-sight, and isotropic (volume-averaged) BAO measurements at an effective redshift $z_{\text{eff}}$ are then sensitive to the transverse angular scale $\theta_d$, redshift span $\delta z_d$, and isotropic angular scale $\theta_v$, which are given by the following:
\begin{eqnarray}
\label{eq:thetad}
\theta_d(z_{\text{eff}}) &=& \frac{r_d}{D_M(z_{\text{eff}})} = \frac{r_dH_0}{\int_0^{z_{\text{eff}}} dz'/E(z')}\,, \\
\label{eq:zd}
\delta z_d(z_{\text{eff}}) &=& \frac{r_d}{D_H(z_{\text{eff}})} = r_dH(z_{\text{eff}})=r_dH_0E(z_{\text{eff}})\,, \\
\label{eq:thetav}
\theta_v(z_{\text{eff}}) &=& \frac{r_d}{D_V(z_{\text{eff}})} = \frac{r_d}{ \left [ z_{\text{eff}} D_M^2(z_{\text{eff}})D_H(z_{\text{eff}}) \right ] ^{1/3}}\,,
\end{eqnarray}
where the sound horizon at baryon drag is determined by the following integral:
\begin{eqnarray}
r_d = \int_{z_d}^{\infty}dz\, \frac{c_s(z)}{H(z)}\,,
\label{eq:rd}
\end{eqnarray}
with $c_s(z)$ being the speed of sound of the photon-baryon plasma, and $z_d \approx 1060$ denoting the redshift of the drag epoch when baryons are released from the photon drag.

A single BAO measurement (at a single effective redshift) of one among $\theta_d$, $\delta z_d$, or $\theta_v$ is unable to disentangle the effects of $r_dH_0$ and $\Omega_m$, which are completely degenerate between each other. However, this degeneracy can be partially broken if the BAO angular scale is measured over a sufficiently wide range of effective redshifts (at present, $0.1 \lesssim z_{\text{eff}} \lesssim 2.5$). The reason is that the slope describing the $r_dH_0$-$\Omega_m$ correlation slowly changes with $z_{\text{eff}}$ (see e.g.\ Fig.~2 of Ref.~\cite{Lin:2021sfs}), reflecting how the importance of the dark energy contribution relative to the matter one decreases with increasing $z_{\text{eff}}$. This is not sufficient to lead to a very constraining inference of $\Omega_m$ at the current level of precision (which is why earlier we stated that the degeneracy is only partially broken), but illustrates where the sensitivity of BAO measurements to $\Omega_m$ arises from.

On the other hand, uncalibrated SNeIa constitute an excellent probe of $\Omega_m$. Neglecting the usual stretch and color corrections, which do not alter our subsequent discussion, we recall that the observed SNeIa light-curve B-band rest-frame peak magnitude $m_B$ is given by:
\begin{eqnarray}
 m_B = M_B - 5\log_{10} \left [ \frac{H_0}{c} \right ] +5\log_{10} \left [ \frac{H_0D_L(z)}{c} \right ] +25\,, \nonumber \\
\label{eq:mb}
\end{eqnarray}
where $M_B$ denotes the SNeIa absolute magnitude in the same band and is treated as a nuisance parameter. In the absence of any knowledge about $M_B$, we refer to the SNeIa measurements as being uncalibrated. These then probe the uncalibrated luminosity distance $H_0D_L(z)$ which, assuming that the Etherington DDR holds, is given by the following:
\begin{eqnarray}
H_0D_L(z_{\text{eff}})&=&H_0(1+z_{\text{eff}})D_M(z) \nonumber \\
&=&(1+z_{\text{eff}})\int_{0}^{z_{\text{eff}}}\, \frac{dz}{E(z)}\,.
\label{eq:dlh0}
\end{eqnarray}
From Eq.~(\ref{eq:dlh0}) it is clear that uncalibrated SNeIa can be used as relative distance indicators to constrain the shape of the late-time expansion rate $E(z)$ regardless of its overall amplitude. Within the minimal $\Lambda$CDM model, and neglecting the radiation component which is completely subdominant at late times, the only free parameter that enters into $E(z) \approx \sqrt{\Omega_m(1+z)^3+(1-\Omega_m)}$ is $\Omega_m$, which can therefore be determined from uncalibrated SNeIa measurements. This determination can be further improved by combining SNeIa with BAO measurements, whose sensitivity to $\Omega_m$ we discussed earlier.

For what concerns the physical baryon density $\omega_b$, this is tightly constrained by BBN considerations on the abundance of light elements, especially deuterium. This is because $\omega_b$ controls (among others) the neutron-to-proton density at the time of BBN, a parameter to which the yield of light elements is particularly sensitive. It is worth noting that the CMB is also sensitive to $\omega_b$ through its impact on the relative height of odd and even acoustic peaks. The signature of $\omega_b$ in CMB data is therefore quite clean, making its determination relatively stable across different models, while still being in part model-dependent. Nevertheless, in support of the BBN determination of $\omega_b$, it is worth noting that this is in excellent agreement with the $\Lambda$CDM-based CMB determination of the same parameter. Moreover, it has been shown in Ref.~\cite{Motloch:2020lhu} that $\omega_b$ can be inferred in eight independent ways from CMB data, reflecting eight independent ways in which the physical baryon density affects the CMB power spectra, and with all eight determinations being in agreement between each other. Therefore, while the earlier calibration of $\Omega_m$ is somewhat model-dependent (it implicitly depends on the assumed late-time background expansion), the calibration of $\omega_b$ can be considered extremely robust and model-independent.

We note that the increase in $\omega_c$ implied by Eqs.~(\ref{eq:deltac}--\ref{eq:deltacplanck2018}) has important implications for the clustering amplitude, as quantified by the parameter $S_8 \equiv \sigma_8\sqrt{\Omega_m/0.3}$, where $\sigma_8$ is the present-day linear theory amplitude of matter fluctuations averaged in spheres of radius $8\,h{\text{Mpc}}^{-1}$. Under the assumption that the primordial power spectrum of scalar fluctuations remains $\Lambda$CDM-like, an increase in $\omega_c$ while maintaining $\omega_b$ fixed as per BBN considerations (and the physical radiation density $\omega_r$ being fixed by the measured CMB temperature monopole) directly leads to an increase in $\omega_m$, and therefore to the onset of matter domination occurring earlier. This, in turn, causes a larger amplitude of fluctuations in the matter power spectrum, as well as a larger net growth of matter perturbations, both of which result in a larger value of $\sigma_8$. Therefore, an increase in $\omega_c$ indirectly leads to an increase in $\sigma_8$. Moreover, if $\Omega_m$ is fixed (note that this fixes the large-scale asymptote of the matter power spectrum, see e.g.\ Fig.~4.6 of Ref.~\cite{Vagnozzi:2019utt}), the small-scale linear matter power spectrum $P_L$ scales with $H_0$ as follows:
\begin{eqnarray}
P_L \sim \frac{T^2(k)}{H_0^4} \sim \frac{k_{\text{eq}}^4}{H_0^4} \sim a_{\text{eq}}^{-2} \sim H_0^4\,,
\label{eq:power}
\end{eqnarray}
where $T(k)$ denotes the transfer function, $k_{\text{eq}}$ is the equality wavenumber, and $a_{\text{eq}}$ is the equality scale factor. This necessarily increases $S_8 \sim H_0^2$ as well, worsening the discrepancy between values of $S_8$ determined by weak lensing~\cite{DES:2017qwj,Heymans:2020gsg}, redshift space distortions~\cite{Efstathiou:2017rgv,Benisty:2020kdt,Nunes:2021ipq}, and cluster counts~\cite{Sakr:2021jya}, all of which fall somewhat lower compared to the \textit{Planck} (TTTEEE+lowE+lensing) determination of $S_8=0.832 \pm 0.013$~\cite{Planck:2018vyg}.~\footnote{We note that recent analyses of cosmic shear data from DES Y3 and KiDS-1000~\cite{Kilo-DegreeSurvey:2023gfr}, and of the cluster mass function from SRG/eROSITA~\cite{Ghirardini:2024yni}, have decreased the significance of the $S_8$ discrepancy.} Analogously to Eq.~(\ref{eq:deltac}), we can expect the following:
\begin{eqnarray}
\frac{\delta S_8}{S_8} \approx 2 \frac{\delta h}{h}\,.
\label{eq:deltas8}
\end{eqnarray}

\section{Datasets and methodology}
\label{sec:data}

Our discussions so far have been purely theoretical. In what follows, we will test on real data the validity of our argument that $H_0$, $\omega_c$, and $S_8$ increase hand-in-hand.

We make use of the following datasets and priors:
\begin{itemize}
\item BAO measurements from the Baryon Oscillation Spectroscopic Survey (BOSS) and extended BOSS (eBOSS) survey programs of the Sloan Digital Sky Survey (SDSS). In particular, the BAO measurements we use come from the Main Galaxy Sample (MGS) at $z_{\text{eff}}=0.15$~\cite{Ross:2014qpa}; the BOSS galaxy samples at $z_{\text{eff}}=0.38$, $0.51$~\cite{BOSS:2016wmc}; the eBOSS LRG sample at $z_{\text{eff}}=0.70$~\cite{eBOSS:2016onl}; the eBOSS ELG sample at $z_{\text{eff}}=0.85$~\cite{eBOSS:2020qek}; the eBOSS QSO sample at $z_{\text{eff}}=1.48$~\cite{eBOSS:2020gbb}; the eBOSS Ly-$\alpha$ sample and the cross-correlation between the Ly-$\alpha$ and QSO samples, both of them at $z_{\text{eff}}=2.33$~\cite{eBOSS:2019ytm}. We refer to this dataset as \textbf{\textit{BAO}}.
\item The \textit{PantheonPlus} SNeIa catalog~\cite{Brout:2022vxf}, consisting of 1701 light curves for 1550 unique SNeIa. We only use SNeIa in the redshift range $0.01<z<2.26$, and refer to this dataset as \textbf{\textit{PP}}.
\item The \textit{Pantheon} Type Ia Supernovae (SNeIa) catalog~\cite{Pan-STARRS1:2017jku}, which precedes the \textit{PantheonPlus} one and consists of 1048 SNeIa within the redshift range $0.01<z<2.26$. We refer to this dataset as \textbf{\textit{P}}.
\item A Gaussian prior on the physical baryon density $\omega_b = 0.02233 \pm 0.00036$, determined from BBN considerations in light of an improved determination of the deuterium burning rate from the LUNA experiment~\cite{Mossa:2020gjc}, and which we refer to as \textbf{\textit{BBN}}.
\item A Gaussian prior on the Hubble constant $H_0=(73.04 \pm 1.04)\,{\text{km}}/{\text{s}}/{\text{Mpc}}$ as determined by the SH0ES team through a Cepheid-calibrated SNeIa distance ladder~\cite{Riess:2021jrx}, and which we refer to as \textbf{\textit{SH0ES}}.
\end{itemize}
We consider various combinations of the above datasets, following a rationale that will be discussed later. Note that we purposely choose not to include CMB data (not even in the form of distance priors), unlike the related Ref.~\cite{Poulin:2024ken}. The reason is that we want our results to depend only on constraints arising from the late-time expansion history as much as possible, whereas including CMB data would introduce an inevitable dependence on the perturbation dynamics of the assumed early-time model, a model-dependence which we are seeking to avoid.

For what concerns the underlying model, we assume that this is $\Lambda$CDM. We note, in addition, that the aforementioned datasets are inherently background ones. Therefore, rather than adopting the usual 6-dimensional parameter basis $\{\theta_s,\omega_b,\omega_c,A_s,n_s,\tau\}$, it makes more sense to work with the 3-dimensional parameter basis $\{\omega_b,\Omega_m,h\}$, since $A_s$, $n_s$, and $\tau$ play no role in what follows (we recall that $\theta_s$ and $H_0$ can be exchanged one for the other). We note that in principle the late-time expansion history is fully specified by $\Omega_m$ and $h$ alone, since it is really only the sum of $\omega_b+\omega_c+\omega_{\nu}=\omega_m=\Omega_mh^2$, rather than the two individual components alone, which matters at the background level. However, we treat $\omega_b$ as an additional free parameter since the key point of our work is to study the relation between $H_0$ and $\omega_c$. This obviously requires disentangling the $\omega_b$ and $\omega_c$ contributions to $\omega_m$, which is achieved through the BBN prior to $\omega_b$ as we will discuss shortly.

We sample the posterior distributions of the three cosmological parameters using Monte Carlo Markov Chain (MCMC) methods, adopting the cosmological MCMC sampler \texttt{SimpleMC}~\footnote{This code was first presented and used in the seminal Ref.~\cite{BOSS:2014hhw} by the BOSS collaboration, and is available at \url{https://github.com/ja-vazquez/SimpleMC}.}. The code can be used to perform parameter estimation against datasets for which only the background expansion history matters, which is precisely the case for the measurements used here. We set wide, flat priors on all three cosmological parameters, verifying a posteriori that our posteriors are not affected by the choice of lower and upper prior boundaries. We assess the convergence of our MCMC chains using the Gelman-Rubin $R-1$ parameter~\cite{Gelman:1992zz}, with $R-1<0.01$ required for our chains to be considered converged. Our chains are subsequently analyzed via the \texttt{GetDist} package~\cite{Lewis:2019xzd}.

Assuming $\Lambda$CDM as the underlying cosmological model, we explicitly investigate on real data the connection between $h$ and $\omega_c$ discussed earlier, see Eq.~(\ref{eq:omegac}). We do so by first calibrating $\Omega_m$ by using BAO and/or SNeIa measurements, as discussed in Sec.~\ref{sec:multidimensionality}. Once this is achieved, we calibrate the resulting distance ladder on opposite ends, resulting in $H_0$ being pushed to either end of the ``Hubble tension range'', $0.67 \lesssim h \lesssim 0.74$. On the one hand, we can use the BBN prior on $\omega_b$ to infer $r_d$ (assuming $\Lambda$CDM), in turn calibrating BAO measurements from the early Universe side (this calibration is eventually transferred to SNeIa measurements in the same range, if they are included). The value of $H_0$ inferred from this inverse distance ladder is expected to fall on the low side of the Hubble tension range, i.e.\ closer to the Planck value~\cite{Schoneberg:2019wmt,Schoneberg:2022ggi}. On the other hand, the SH0ES prior on $H_0$ can be used to directly calibrate either or both BAO and SNeIa measurements from the local Universe side, naturally resulting in a value of $H_0$ falling on the high side of the Hubble tension range, i.e.\ closer to the SH0ES value.

As the two calibrations discussed above move $H_0$ across the Hubble tension range, we can expect $\omega_c$ to increase/decrease as $H_0$ does the same. Two clarifications are in order before moving on. Firstly, as we will discuss in more detail in Sec.~\ref{sec:results}, at times we will consider dataset combinations including SNeIa but not BAO, yet we will still include the BBN prior on $\omega_b$. In these cases, the role of the BBN prior is to disentangle the $\omega_b$ and $\omega_c$ contributions to $\omega_m=\omega_b+\omega_c+\omega_{\nu}$, which are otherwise completely degenerate as far as SNeIa data is concerned. Note that when combined with BAO measurements, the BBN prior still plays this role of disentangling $\omega_b$ and $\omega_c$, in addition of course to calibrating $r_d$. Finally we note that, while the value of $H_0$ inferred from the BAO+BBN combination inherently depends on the assumed early Universe model, required to infer $r_d$ from $\omega_b$ via Eq.~(\ref{eq:rd}), the relation between $H_0$ and $\omega_c$ is largely independent of the assumed early Universe physics. In fact, such a relation is implied not by the assumed early-Universe physics, but by the constraints on $\Omega_m$ imposed from the late-time expansion history (see also Ref.~\cite{Poulin:2024ken} for further discussions on this point).

In addition, we also study the connection between $h$ and $S_8$ discussed earlier, see Eq.~(\ref{eq:deltas8}). To do so, we compute $S_8$ (obviously treated as a derived parameter) for each point in our MCMC chains, using the Boltzmann solver \texttt{CAMB}~\cite{Lewis:1999bs}. As already emphasized earlier, this calculation requires explicit knowledge of the primordial power spectrum of scalar fluctuations, which we assume remains $\Lambda$CDM-like (see also Ref.~\cite{Poulin:2024ken} where a similar assumption was made). We assume that the amplitude of the primordial power spectrum is fixed by the value of $A_s$ inferred by a fit to the \textit{Planck} 2018 TTTEEE+lowE+lensing likelihoods. Nevertheless, insofar as we only care about shifts in $S_8$ (in response to shifts in $H_0$) rather than the reference value thereof, as is the case in this work, we do not expect our subsequent results to depend on the assumed value of $A_s$.

Finally, for purely illustrative purposes, we further quantify the relation between $\omega_c$, $S_8$, and $H_0$ [see Eqs.~(\ref{eq:deltaomegacdeltah},\ref{eq:deltas8})] by fitting linear relations to the relative variations of both $\omega_c$ and $S_8$ ($\delta \omega_c/\omega_c$ and $\delta S_8/S_8$) as a function of the relative variation of $H_0$, $\delta h/h$:
\begin{eqnarray}
\frac{\delta\omega_c}{\omega_c} = \alpha \frac{\delta h}{h}\,, \quad \frac{\delta S_8}{S_8} = \beta \frac{\delta h}{h}\,.
\label{eq:alphabeta}
\end{eqnarray}
For each dataset combination, we determine the best-fit coefficients $\alpha$ and $\beta$ from a least-squares fit, minimizing the sum of the squares of the residuals, $\sum \left ( \alpha \delta h/h - \delta \omega_c/\omega_c \right ) ^2$ and $\sum \left ( \beta \delta h/h - \delta S_8/S_8 \right ) ^2$ respectively. For each set of chains, denoting by $x$ a given parameter among $\{h,\omega_c,S_8\}$, and by $\bar{x}$ its mean value (computed across the chains), we estimate $\delta x/x$ as $\delta x/x=2(x - \bar{x})/(x+\bar{x})$.

\section{Results}
\label{sec:results}

We now present the results obtained using the methods and datasets discussed in Sec.~\ref{sec:data}. Constraints on cosmological parameters of interest, as well as the coefficients $\alpha$ and $\beta$ presented in Eq.~(\ref{eq:alphabeta}), are shown in Tab.~\ref{tab:results}. The following four subsections are each devoted to the comparison between the results obtained adopting two specific combinations of likelihoods discussed previously, each of which will make the mutual correlations between $\omega_c$, $S_8$, and $H_0$ more or less clear, while allowing us to underscore the importance of a reliable $\Omega_m$ calibration.

\subsection{BBN+PP+BAO vs BBN+PP+BAO+SH0ES}
\label{subsec:bbnppbaobbnppbaosh0es}

\begin{figure}[!ht]
\centering
\includegraphics[width=1\linewidth]{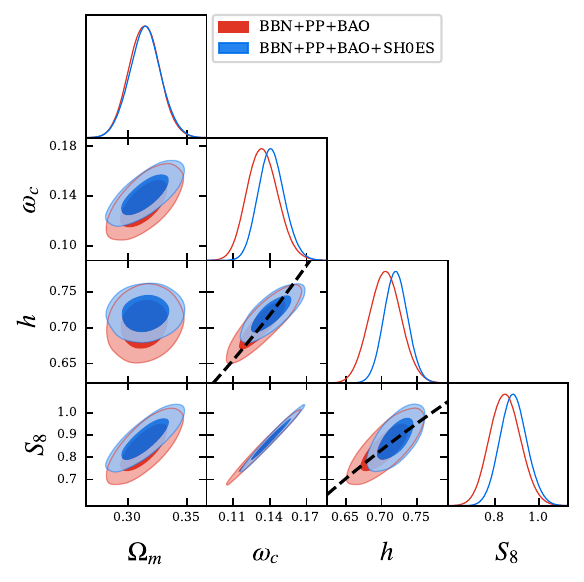}
\caption{Triangular plot showing 2D joint and 1D marginalized posterior probability distributions for the fractional matter density parameter $\Omega_m$, the physical cold dark matter density $\omega_c$, the reduced Hubble constant $h$, and the clustering parameter $S_8$, in light of the \textit{BBN}+\textit{PP}+\textit{BAO} (red contours and curves) and \textit{BBN}+\textit{PP}+\textit{BAO}+\textit{SH0ES} (blue contours and curves) dataset combinations. We clearly see that $h$, $\omega_c$, and $S_8$ increase hand-in-hand. The dashed lines shows the linearized theoretical estimates for the relationships between $\delta \omega_c/\omega_c$ and $\delta h/h$ given by Eq.~(\ref{eq:deltacplanck2018}), and between $\delta S_8/S_8$ and $\delta h/h$ given by Eq.~(\ref{eq:deltas8}).}
\label{fig:wwosh0es}
\end{figure}

We begin by comparing the results obtained from the \textit{BBN}+\textit{PP}+\textit{BAO} versus \textit{BBN}+\textit{PP}+\textit{BAO}+\textit{SH0ES} dataset combinations. A visual summary of our results is given in the corner plot of Fig.~\ref{fig:wwosh0es}, where we show 2D joint and 1D marginalized posterior probability distributions for $\Omega_m$, $\omega_c$, $h$, and $S_8$.

The rationale behind this first comparison is that the \textit{BAO}+\textit{PP} combination is able to calibrate $\Omega_m$ (as we discussed in Sec.~\ref{sec:multidimensionality}) in a consistent way across both dataset combinations. Indeed, for both dataset combinations we find $\Omega_m = 0.314 \pm 0.013$, see Tab.~\ref{tab:results}. On the other hand, as discussed in Sec.~\ref{sec:data}, calibrating the \textit{BAO}+\textit{PP} dataset combination with either the \textit{BBN} prior on $\omega_b$ or the \textit{SH0ES} prior on $H_0$ pushes $H_0$ to opposite ends, allowing us to explore how $\omega_c$ and $S_8$ change in response to these different calibrations. We note that from the \textit{BBN}+\textit{PP}+\textit{BAO} versus \textit{BBN}+\textit{PP}+\textit{BAO}+\textit{SH0ES} dataset combinations we infer $H_0=(70.6 \pm 2.2)\,{\text{km}}/{\text{s}}/{\text{Mpc}}$ and $H_0=(72.0 \pm 1.7)\,{\text{km}}/{\text{s}}/{\text{Mpc}}$ respectively.

As we can clearly see in the triangular plot of Fig.~\ref{fig:wwosh0es}, the expected mutual correlations between $h$, $\omega_c$, and $S_8$ are clearly present, although the shift in $\omega_c$ and $S_8$ as the calibration is changed is not strong (less than $1\sigma$ in both cases). We find that $\omega_c$ increases from $0.134 \pm 0.013$ to $0.141 \pm 0.011$, whereas $S_8$ increases from $0.846 \pm 0.071$ to $0.882 \pm 0.062$. The reason for these relatively small shifts is due to the fact that the shift in $H_0$ itself is not large. As one would expect, adding \textit{BBN}+\textit{BAO} (which in itself favors lower values of $H_0$) to the \textit{PP}+\textit{SH0ES} combination, which on its own would favor values of $H_0$ closer to $74\,{\text{km}}/{\text{s}}/{\text{Mpc}}$, brings this value down to $72.0\,{\text{km}}/{\text{s}}/{\text{Mpc}}$. For the \textit{BBN}+\textit{PP}+\textit{BAO} dataset combination our best-fit values of $\alpha$ and $\beta$ are $\alpha=2.54$ and $\beta=2.06$, whereas for the \textit{BBN}+\textit{PP}+\textit{BAO}+\textit{SH0ES} combination we find $\alpha=2.49$ and $\beta=1.98$, both aligning relatively well with the analytical arguments presented earlier (represented by the dashed black lines in Fig.~\ref{fig:wwosh0es}).

\subsection{BBN+PP+BAO vs BBN+PP+SH0ES}
\label{subsec:bbnppbaobbnppsh0es}

\begin{figure}[!ht]
\centering
\includegraphics[width=1\linewidth]{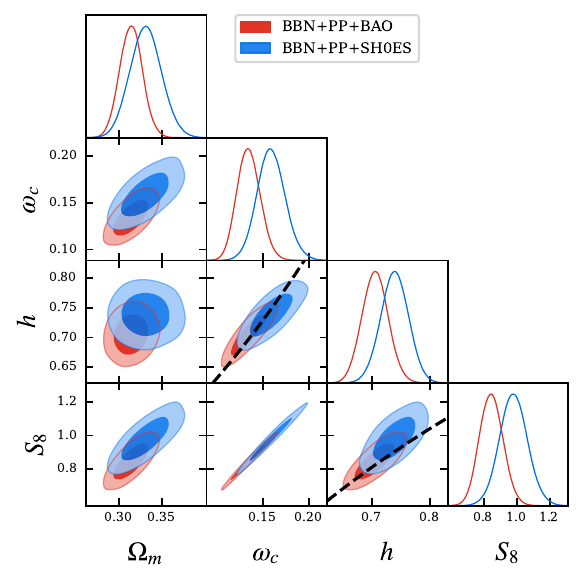}
\caption{As in Fig.~\ref{fig:wwosh0es}, but focusing on the \textit{BBN}+\textit{PP}+\textit{BAO} (red contours and curves) versus \textit{BBN}+\textit{PP}+\textit{SH0ES} (blue contours and curves) dataset combinations.}
\label{fig:baovssh0es}
\end{figure}

We now move on to comparing the results obtained from the \textit{BBN}+\textit{PP}+\textit{BAO} versus \textit{BBN}+\textit{PP}+\textit{SH0ES} dataset combinations. In practice, compared to the earlier discussion in Sec.~\ref{subsec:bbnppbaobbnppbaosh0es}, we have removed the \textit{BAO} dataset from the second combination. A visual summary of our results is given in the corner plot of Fig.~\ref{fig:baovssh0es}.

In this case, we are no longer adopting the same dataset to calibrate $\Omega_m$ (earlier constrained in both cases by the \textit{BAO}+\textit{PP} combination), which is calibrated by \textit{BAO}+\textit{PP} on one side, and by \textit{PP} on the other. The resulting calibration of $\Omega_m$ is broadly consistent, but displays some differences. In particular, from the \textit{BBN}+\textit{PP}+\textit{BAO} combination we find $\Omega_m=0.314 \pm 0.013$, as previously discussed in Sec.~\ref{subsec:bbnppbaobbnppbaosh0es}, whereas from the \textit{BBN}+\textit{PP}+\textit{SH0ES} combination we find the $\approx 0.8\sigma$ larger value of $\Omega_m=0.331 \pm 0.018$, as can be clearly appreciated in Fig.~\ref{fig:baovssh0es}. This is not unexpected, as the fact that the \textit{PantheonPlus} SNeIa sample appears to prefer slightly larger values of $\Omega_m \sim 0.33$ compared to its predecessors \textit{Pantheon} and \textit{JLA} is well known and documented in the literature~\cite{Baryakhtar:2024rky}.

Comparing these two dataset combinations we find that the expected mutual correlations between $h$, $\omega_c$, and $S_8$ are again clearly present, but are accentuated compared to the earlier results of Sec.~\ref{subsec:bbnppbaobbnppbaosh0es}. The reason is two-fold. On the one hand, from entirely analytical arguments [see Eq.~(\ref{eq:omegam})], one expects that if the calibration of $\Omega_m$ is not constant, larger values thereof (as in the \textit{BBN}+\textit{PP}+\textit{SH0ES} combination) will correlate with larger values of $\omega_c$ and $S_8$, which is precisely what we are observing. In addition, removing the \textit{BAO} dataset from the \textit{BBN}+\textit{PP}+\textit{SH0ES} combination allows the latter to push $H_0$ towards $74\,{\text{km}}/{\text{s}}/{\text{Mpc}}$ as expected based on our earlier discussions. Indeed, we find $H_0=(73.9 \pm 2.5)\,{\text{km}}/{\text{s}}/{\text{Mpc}}$, completely in line with what one would expect.

The above increase in $H_0$ leads to substantial increases in both $\omega_c$ and $S_8$. In fact, we find that $\omega_c$ increases by $1.3\sigma$ from $0.134 \pm 0.013$ to $0.159 \pm 0.015$: we remark that the presence of the \textit{BBN} prior in the \textit{BBN}+\textit{PP}+\textit{SH0ES} dataset combination plays a crucial role in determining the latter value, given that \textit{PP}+\textit{SH0ES} on their own would only be sensitive to the sum of $\omega_b$ and $\omega_c$ (more precisely, \textit{PP} is sensitive to $\Omega_m$, so adding the \textit{SH0ES} information on $H_0$ naturally determines $\omega_m$), whereas the \textit{BBN} prior allows to disentangle the baryonic contribution from the DM one. Similarly to $\omega_c$, we observe that $S_8$ increases from $0.846 \pm 0.071$ to $0.980 \pm 0.084$. In the case of $S_8$, although the central value clearly increases to extremely high values, it would be misleading to quantify the significance of the resulting tension with either CMB or weak lensing observations. In fact, we note that the uncertainty on $S_8$ is of order $0.07$-$0.08$, i.e.\ a factor of $\approx 6$-$7$ larger than that of $S_8=0.832 \pm 0.013$ as determined by \textit{Planck}, which would result in a tension formally of low significance ($\lesssim 2\sigma$) in spite of the very high central value. The reason for these large uncertainties is to be sought in the fact that our analysis includes neither CMB nor weak lensing data, both of which are crucial to reduce the uncertainty on $S_8$. Nevertheless, even in this way we are able to observe a significant upwards shift in $S_8$, in line with the analytical expectations laid out previously. Finally, the best-fit values for $\alpha$ and $\beta$ we determine for the \textit{BBN}+\textit{PP}+\textit{SH0ES} combination are $\alpha=2.27$ and $\beta=1.73$.

\subsection{BBN+BAO vs BBN+PP+SH0ES}
\label{subsec:bbnbaobbnppsh0es}

\begin{figure}[!ht]
\centering
\includegraphics[width=1\linewidth]{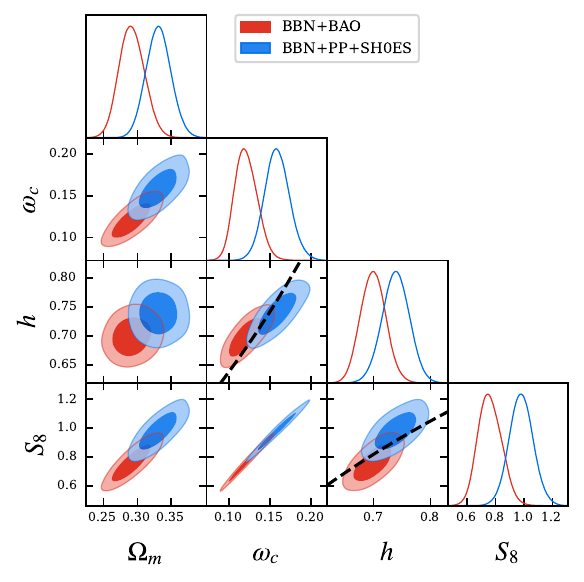}
\caption{As in Fig.~\ref{fig:wwosh0es}, but focusing on the \textit{BBN}+\textit{BAO} (red contours and curves) versus \textit{BBN}+\textit{PP}+\textit{SH0ES} (blue contours and curves) dataset combinations.}
\label{fig:baovsppsh0es}
\end{figure}

Driven by the previous results, we now choose to compare two dataset combinations which exacerbate the tension between the respective determinations of $H_0$, in order to check whether a larger shift is observed in both $\omega_c$ and $S_8$. Specifically, we compare the \textit{BBN}+\textit{BAO} versus \textit{BBN}+\textit{PP}+\textit{SH0ES} dataset combinations: compared to the earlier discussion in Sec.~\ref{subsec:bbnppbaobbnppsh0es}, we have removed the \textit{PP} dataset from the first combination. A visual summary of our results is given in the corner plot of Fig.~\ref{fig:baovsppsh0es}.

In this case, analogously to Sec.~\ref{subsec:bbnbaobbnppsh0es}, we are no longer adopting the same dataset to calibrate $\Omega_m$, which is calibrated by the \textit{BAO} dataset on one side, and the \textit{PP} dataset on the other. In terms of the resulting value of $\Omega_m$, these two datasets display the largest difference among all the ones we discussed. In particular, from the \textit{BBN}+\textit{BAO} combination we find $\Omega_m=0.292 \pm 0.019$, whereas from the \textit{BBN}+\textit{PP}+\textit{SH0ES} combination we find the $\approx 1.5\sigma$ larger value of $\Omega_m=0.331 \pm 0.018$ discussed previously in Sec.~\ref{subsec:bbnppbaobbnppsh0es}.

Comparing these two dataset combinations leads to the largest differences between the inferred values of $H_0$, for which we find $H_0=(69.9 \pm 2.2)\,{\text{km}}/{\text{s}}/{\text{Mpc}}$ from \textit{BBN}+\textit{BAO}, and $H_0=(73.9 \pm 2.5)\,{\text{km}}/{\text{s}}/{\text{Mpc}}$ from \textit{BBN}+\textit{PP}+\textit{SH0ES} as discussed earlier. Again in line with expectations, we find correspondingly large shifts in both $\omega_c$, which increases by $1.9\sigma$ from $0.121 \pm 0.014$ to $0.159 \pm 0.015$, as well as $S_8$ which increases from $0.760 \pm 0.084$ to $0.980 \pm 0.084$ (again, we refrain from quoting the level of significance of the $S_8$ tension due to the large uncertainties).

The significant shifts in $\omega_c$ and $S_8$ reported above, while going precisely in the direction expected from our analytical considerations, are however exacerbated by the inconsistent calibration of $\Omega_m$. This is very clear from Fig.~\ref{fig:baovsppsh0es}, which shows that the larger value of $\Omega_m$ in the \textit{BBN}+\textit{PP}+\textit{SH0ES} case is partially responsible for driving $\omega_c$ and $S_8$ towards even larger values, as one can expect from the analytical argument presented in Eq.~(\ref{eq:omegam}). These findings underscore the capital importance of a reliable calibration of $\Omega_m$, as even $1.5\sigma$ shifts in this parameter are sufficient to drive significant shifts in $\omega_c$ and $S_8$. Finally, for the \textit{BAO}+\textit{BBN} combination we find the best-fit values $\alpha=2.70$ and $\beta=2.27$.

\subsection{BBN+BAO vs BBN+P+SH0ES}
\label{subsec:bbnbaobbnpsh0es}

\begin{figure}[!ht]
\centering
\includegraphics[width=1\linewidth]{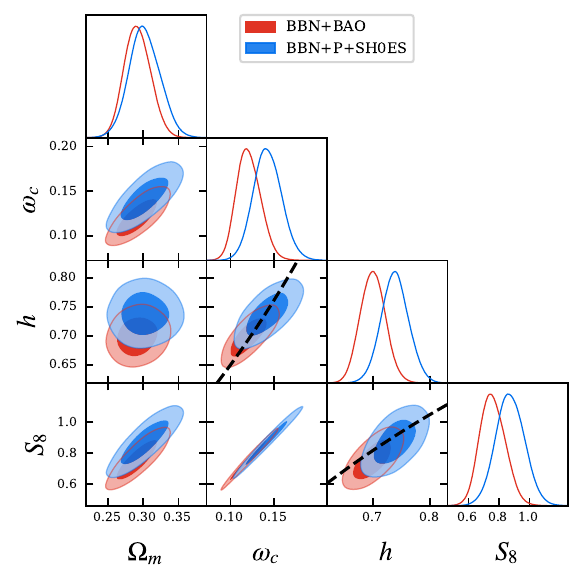}
\caption{As in Fig.~\ref{fig:wwosh0es}, but focusing on the \textit{BBN}+\textit{BAO} (red contours and curves) versus \textit{BBN}+\textit{P}+\textit{SH0ES} (blue contours and curves) dataset combinations.}
\label{fig:baovspsh0es}
\end{figure}

In light of all the previous considerations, the final scenario we study is one where $\Omega_m$ is still calibrated by two different probes, but in a way which is more consistent than the ones considered in Sec.~\ref{subsec:bbnppbaobbnppsh0es} and Sec.~\ref{subsec:bbnbaobbnppsh0es}. In both cases, we saw that the use of the \textit{PantheonPlus} SNeIa catalog was responsible for the higher value of $\Omega_m \sim 0.33$ inferred in the respective dataset combination. However, it is known that preceding \textit{Pantheon} SNeIa catalog preferred lower values of $\Omega_m$ compared to \textit{PantheonPlus}. For this reason, we now choose to compare the \textit{BBN}+\textit{BAO} versus \textit{BBN}+\textit{P}+\textit{SH0ES} dataset combinations: compared to the earlier discussion in Sec.~\ref{subsec:bbnbaobbnppsh0es}, we have therefore replaced the \textit{PP} dataset in the second combination with the \textit{P} one. A visual summary of our results is given in the corner plot of Fig.~\ref{fig:baovspsh0es}. In some sense, this comparison is the best compromise among all the ones we have considered so far. In fact, such a comparison allows for a highly consistent calibration of $\Omega_m$, while allowing us to explore the opposite ends of the Hubble tension range precisely because the calibration of $\Omega_m$ is achieved via different datasets (BAO and SNeIa), which nonetheless are consistent with each other as far as $\Omega_m$ is concerned.

As expected, we find that $\Omega_m$ is consistently calibrated across the two dataset combination. From \textit{BBN}+\textit{BAO} we infer $\Omega_m = 0.292 \pm 0.019$ as already discussed earlier, whereas from \textit{BBN}+\textit{P}+\textit{SH0ES} we find $\Omega_m=0.301 \pm 0.022$, which is consistent within $0.3\sigma$. At the same time, the \textit{BBN} calibration on one side and the \textit{SH0ES} calibration on the other push $H_0$ to opposite ends of the Hubble tension range, with $H_0=(69.9 \pm 2.2)\,{\text{km}}/{\text{s}}/{\text{Mpc}}$ from the \textit{BBN}+\textit{BAO} dataset combination as already discussed earlier, and $H_0=(73.8 \pm 2.4)\,{\text{km}}/{\text{s}}/{\text{Mpc}}$ from the \textit{BBN}+\textit{P}+\textit{SH0ES} one. In line with our expectations, we find that $\omega_c$ increases by $1\sigma$ from $0.121 \pm 0.014$ to $0.142 \pm 0.016$, whereas $S_8$ which increases from $0.760 \pm 0.084$ to $0.871 \pm 0.094$. The shifts are smaller compared to those reported in Sec.~\ref{subsec:bbnbaobbnppsh0es}, and the reason is precisely because $\Omega_m$ is consistently calibrated across the two datasets, albeit using different probes. Such a consistent calibration removes any spurious shift in $\omega_c$ and $S_8$ resulting from an increase in the underlying value of $\Omega_m$, underscoring once more the importance of a reliable calibration of the latter. We will return to this point shortly. Finally, for the \textit{BAO}+\textit{P}+\textit{SH0ES} combination we find the best-fit values $\alpha=2.33$ and $\beta=1.84$.

\begin{table*}[!htb]
\footnotesize
\scalebox{1.2}{
\begin{tabular}{|c?c|c|c|c|c|c|}
\hline
\textbf{Dataset combination} & $\Omega_m$ & $H_0\,[{\text{km}}/{\text{s}}/{\text{Mpc}}]$ & $\omega_c$ & $S_8$ & $\alpha$ & $\beta$ \\
\hline\hline
\textit{BBN}+\textit{BAO} & $0.292 \pm 0.019$ & $69.9 \pm 2.2$ & $0.121 \pm 0.014$ & $0.760 \pm 0.084$ & 2.70 & 2.27 \\ \hline\
\textit{BBN}+\textit{PP}+\textit{SH0ES} & $0.331 \pm 0.018$ & $73.9 \pm 2.5$ & $0.159 \pm 0.015$ & $0.980 \pm 0.084$ & 2.27 & 1.73 \\ \hline
\textit{BBN}+\textit{PP}+\textit{BAO}+\textit{SH0ES} & $0.314 \pm 0.013$ & $72.0 \pm 1.7$ & $0.141 \pm 0.011$ & $0.882 \pm 0.062$ & 2.49 & 1.98 \\ \hline
\textit{BBN}+\textit{P}+\textit{SH0ES} & $0.301 \pm 0.022$ & $73.8 \pm 2.4$ & $0.142 \pm 0.016$ & $0.871 \pm 0.094$ & 2.33 & 1.84 \\ \hline
\textit{BBN}+\textit{PP}+\textit{BAO} & $0.314 \pm 0.013$ & $70.6 \pm 2.2$ & $0.134 \pm 0.013$ & $0.846 \pm 0.071$ & 2.54 & 2.06 \\ \hline
\end{tabular}}
\caption{68\% credible intervals on the fractional matter density parameter $\Omega_m$, the physical cold dark matter density $\omega_c$, the Hubble constant $H_0$, the clustering parameter $S_8$, and the parameters $\alpha$ and $\beta$ introduced in Eq.~(\ref{eq:alphabeta}), in light of various dataset combinations, all of which are discussed in Sec.~\ref{sec:results}.}
\label{tab:results}
\end{table*}

\section{Discussion}
\label{sec:discussion}

The results presented previously fully confirm on real data the expected correlations between $H_0$, $\omega_c$, and $S_8$ which, we recall, increase hand-in-hand. While we have specifically assumed $\Lambda$CDM at early times, the increase in $\omega_c$ and $S_8$ as $H_0$ increases does not depend on the specific model assumed for the early Universe, since it is solely a consequence of constraints on $\Omega_m$ imposed from the late-time expansion history~\footnote{In other words, as noted in Ref.~\cite{Poulin:2024ken}, uncalibrated late-time data can constrain ``dimensionless'' quantities such as $\Omega_m$, whereas a calibration is required to constrain ``dimensionful'' quantities such as $\omega_c$ -- the terms dimensionless and dimensionful here are used with a slight abuse of language}. In fact, repeating our analysis in a early Universe-agnostic manner, treating $r_d$ as a free parameter and eventually imposing some prior thereon to calibrate the BAO measurements from the early Universe side, would lead to essentially the same conclusions. Therefore, we can expect that any early-time modification to $\Lambda$CDM aiming to solve the Hubble tension will necessarily have to be accompanied by an increase in $\omega_c$ and, if the primordial power spectrum is $\Lambda$CDM-like, in $S_8$ as well. We also note that we have explicitly assumed that $\Lambda$CDM holds at late times, as is usually done when studying early-time models of new physics -- we return to this point the end of the Section.

One point worthy of notice is that the increase in $\omega_c$ is actually welcome from the perspective of early-time models. As explicitly argued by one of us in Ref.~\cite{Vagnozzi:2021gjh}, many such models, especially those which increase the pre-recombination expansion rate, inevitably lead to an enhanced early integrated Sachs-Wolfe (eISW) effect, as a consequence of their contribution to the decay of gravitational potentials. While some of these models possess ingredients which can allow them to balance the extra eISW power (see e.g.\ Refs.~\cite{Niedermann:2020qbw,Escudero:2021rfi}), most of the others do so precisely through an increase in $\omega_c$. The reason is that such an increase anticipates the onset of matter domination, therefore reducing the period over which gravitational potentials decay and drive the eISW effect. In some sense, this increase in $\omega_c$ kills two birds with one stone: it allows early-time models to fit constraints on $\Omega_m$ imposed from the late-time expansion history, and brings the eISW power back to a level which is in good agreement with CMB data.

A very important point which is underscored by our analysis is the key role played by a (consistent or not) $\Omega_m$ calibration. As highlighted especially in Sec.~\ref{subsec:bbnppbaobbnppsh0es} and Sec.~\ref{subsec:bbnbaobbnppsh0es}, an inconsistent calibration of $\Omega_m$ enhances the shifts in $\omega_c$ discussed in this work, which in turn exacerbates the $S_8$ discrepancy. This issue is actually highly relevant at present, given the mild disagreement between the values of $\Omega_m$ inferred from SNeIa catalogs: as we have discussed earlier, the \textit{Pantheon} SNeIa dataset indicated $\Omega_m \sim 0.3$, which increased to $\Omega_m \sim 0.33$ in the \textit{PantheonPlus} sample. The more recent DES-Y5 and Union3 SNeIa samples instead indicate $\Omega_m \sim 0.35$~\cite{DES:2024tys} and $\Omega_m \sim 0.36$~\cite{Rubin:2023ovl} respectively, see also Refs.~\cite{Baryakhtar:2024rky,Colgain:2024ksa} (note that these figures are obtained within $\Lambda$CDM, and potentially increase to much larger values when allowing for an evolving dark energy component): we therefore expect that adopting these SNeIa samples would exacerbate the $\omega_c$ and $S_8$ shifts we have observed. Finally, while we defer a more detailed exploration of this point elsewhere, we note that these shifts in the values of $\Omega_m$ from different samples may be connected to recent discussions on redshift-evolution of inferred cosmological parameters as different redshift are probed (see e.g.\ Refs.~\cite{Krishnan:2020obg,Krishnan:2020vaf,Dainotti:2021pqg,Dainotti:2022bzg,Colgain:2022nlb,Dainotti:2022rea,Colgain:2022rxy,Colgain:2022tql,Jia:2022ycc,Malekjani:2023ple,Adil:2023jtu,Akarsu:2024qiq,Colgain:2024xqj,Colgain:2024clf,Jia:2024wix} for more details).

The above discussions reinforce the urgent need, in the cosmology community, for a calibration of $\Omega_m$ which is as reliable and model-independent as possible. Concerning this last point, we indeed note that the value of $\Omega_m$ inferred from standard late-time cosmological probes (e.g.\ BAO, SNeIa, cosmic chronometers, and so on) is inherently dependent on the assumed (late-time) cosmological model, and will in generally change if one changes the dark energy dynamics. One interesting possibility towards a model-independent determination of $\Omega_m$ could come from measurements of the gas mass fraction $f_{\text{gas}}$ in relaxed, massive galaxy clusters, where $\approx 92$-$95\%$ of the baryons are hot and emit strongly in X-rays. This method relies only on the assumption that the matter content of rich galaxy clusters provides a fair sample of the matter content of the Universe or, in practice, that the cluster potential is able to retain all the matter within the comoving virial radius from which it formed: under these assumptions, the observed ratio of baryonic to total mass should be equal to $\Omega_b/\Omega_m$, independently of any assumed cosmological model, and such a method has indeed been used over the past decades to infer $\Omega_m$~\cite{Allen:2002sr,Ettori:2002pe,Ettori:2009wp,Mantz:2014xba,SPT:2021vsu}.~\footnote{In practice, this method requires making implicit assumptions on General Relativity being the underlying theory of gravity when extrapolating certain scaling relations, but the dependence on the assumed model of gravity is weak.} While the $f_{\text{gas}}$ method is not as widely used in the cosmological community as other probes, we believe that its minimal sensitivity to cosmological model assumptions and relatively high level of maturity compared to other less used probes should make it a very important player in the quest towards solving the Hubble tension, in the interest of calibrating $\Omega_m$ as model-independently as possible. Another interesting possibility towards model-independently inferring $\Omega_m$ (with $\Omega_b$ known) is to use the internal properties of individual galaxies, such as stellar mass, stellar metallicity, and maximum circular velocity~\cite{Villaescusa-Navarro:2022twv,CAMELS:2023ywa,Chawak:2023bil,Hahn:2023upe,Wang:2024dvm}. Nevertheless, this very interesting method, which demonstrates a potential very tight link between cosmological parameters and the astrophysics of galaxies, has yet to reach the level of maturity of all the other probes discussed so far.

In line with the previous comments, we note that our results do depend on the assumed late-time model, which here we have taken to be $\Lambda$CDM, as is usually done when studying early-time models of new physics. In line with the recent findings of Ref.~\cite{Poulin:2024ken} (see also Ref.~\cite{Jedamzik:2020zmd}), we can therefore conclude that successful early-time new physics must be able to reduce $r_d$ while either reducing $\Omega_m$ or increasing $\omega_m$ (a related result was obtained in Ref.~\cite{Jedamzik:2020zmd}, which however relied on a completely different argument, exploiting the different $r_d$-$H_0$ degeneracy directions for BAO versus geometrical information from the CMB). In this sense, we agree with Ref.~\cite{Poulin:2024ken} that, when combined with early-time new physics which reduces $r_d$, the role of late-time new physics can be that of helping relax the constraints on either or both $\Omega_m$ and $S_8$, for instance by allowing for more freedom in the dark energy sector. We note that achieving a successful early-plus-late combination is still not a trivial feat, as not all early-time modifications follow the ideal degeneracy directions.~\footnote{See for instance the explicit example in Ref.~\cite{Toda:2024ncp}, which fails precisely because of the varying electron mass model not following the ideal degeneracy direction for what concerns $\Omega_m$, as also emphasized in Ref.~\cite{Poulin:2024ken} (see also Refs.~\cite{Baryakhtar:2024rky,Schoneberg:2024ynd}). Other recent case studies combining early- and late-time new physics can be found e.g.\ in Refs.~\cite{Allali:2021azp,Ye:2021iwa,Anchordoqui:2021gji,Khosravi:2021csn,Wang:2022jpo,Anchordoqui:2022gmw,Reeves:2022aoi,daCosta:2023mow,Wang:2024dka,Baryakhtar:2024rky}.} However, as also emphasized in Ref.~\cite{Poulin:2024ken} (see also Ref.~\cite{Baryakhtar:2024rky}), this task is potentially made much easier by the recent DESI BAO data which, in order to be as conservative as possible, we have not adopted here.~\footnote{However, we note that even when adopting DESI BAO data deviations from $\Lambda$CDM in the shape of the late-time expansion history remain constrained to $\lesssim 10\%$, as shown in the recent non-parametric analysis of Ref.~\cite{Jiang:2024xnu} by some of us.} Moreover, we remark that ``dark scattering'' models which feature pure momentum exchange, and do not alter the background to linear order in perturbations, can be particularly promising in terms of relaxing the constraints on $\sigma_8$ and, potentially, $\Omega_m$ (see e.g.\ Refs.~\cite{Simpson:2010vh,Escudero:2015yka,Koivisto:2015qua,Pourtsidou:2016ico,Kumar:2017bpv,Gluscevic:2017ywp,Boddy:2018kfv,Boddy:2018wzy,Asghari:2019qld,Vagnozzi:2019kvw,Jimenez:2020ysu,Becker:2020hzj,Figueruelo:2021elm,Vagnozzi:2021quy,Jimenez:2021ybe,Linton:2021cgd,Rogers:2021byl,Ferlito:2022mok,Poulin:2022sgp,Jimenez:2022evh}). We believe it may be worth exploring these models in combination with successful early-time modifications, as also suggested in Ref.~\cite{Vagnozzi:2023nrq}.~\footnote{Of course, as discussed in Ref.~\cite{Vagnozzi:2023nrq}, there is the possibility that additional very-late-time or local new physics may play an important role in the Hubble tension, see e.g.\ Refs.~\cite{Desmond:2019ygn,Ding:2019mmw,Benevento:2020fev,Desmond:2020wep,Cai:2020tpy,Camarena:2021jlr,Cai:2021wgv,Marra:2021fvf,Krishnan:2021dyb,Perivolaropoulos:2021bds,Castello:2021uad,Camarena:2022iae,Perivolaropoulos:2022khd,Oikonomou:2022tjm,Hogas:2023vim,Hogas:2023pjz,Giani:2023aor,Mazurenko:2023sex,Huang:2024erq,Ruchika:2024ymt} for examples of studies in this direction.}

We close with a few remarks. Firstly, our work provides an unified explanation for the $H_0$-$S_8$ correlations frequently observed in the literature on the Hubble tension, but typically explained on a model-by-model basis. We note that another unified explanation for these correlations was provided in Ref.~\cite{Jedamzik:2020zmd}: however, the latter relied on a completely different argument compared to ours, and more specifically was based on the different slopes of the $r_d$-$H_0$ correlation in BAO and (geometrical) CMB data. Our explanation for this correlation is instead based on the (arguably simpler) Eqs.~(\ref{eq:omegam},\ref{eq:omegac}): to the best of our knowledge, it is the first time that this argument is being clearly presented in the literature. Moreover, while for simplicity we have assumed the $\Lambda$CDM model at late times, we remark that our argument is completely general and holds for any model where Eqs.~(\ref{eq:omegam},\ref{eq:omegac}) are valid, while $\Omega_m$ and $\omega_b$ are consistently calibrated. A notable exception could be one where the spatial curvature of the Universe is non-zero, and therefore the contribution of the spatial curvature parameter $\Omega_K$ should be included in the previous equalities. Indeed, our entire discussion, especially the analytical arguments in Sec.~\ref{sec:multidimensionality}, have explicitly assumed a spatially flat Universe. Relaxing this assumption is likely to relax all our results, given the well-known degeneracies between $\Omega_m$ and $\Omega_K$, and we plan to explore this in future work.

\section{Conclusions}
\label{sec:conclusions}

In spite of its name, it is now very clear that the Hubble tension has important implications for other cosmological quantities as well, and that just focusing on $H_0$ is misleading. The sound horizon at baryon drag $r_d$ is another important quantity at play but, even then, looking at $H_0$ and $r_d$ alone obscures part of the story. In this work, we have demonstrated that $\Omega_m$ and $\omega_c$ play a particularly important role in this context.

We have argued that, if both $\Omega_m$ and $\omega_b$ are calibrated (respectively by BAO and/or uncalibrated SNeIa, and by BBN considerations), then an increase in $H_0$ \textit{must} necessarily be accompanied by an increase in $\omega_c$. Under the assumption that the primordial power spectrum of scalar fluctuations remains $\Lambda$CDM-like, an increase in $\omega_c$, via the associated increase in $\omega_m$ (since $\omega_b$ is calibrated), implies an increase in the clustering parameter $S_8$, worsening the mild $S_8$ discrepancy. These shifts are not a consequence of the effects of new physics at early times, but follow solely from late-time expansion history constraints. Therefore, successful early-time new physics models must be able to accommodate these changes (which, in the case of $\omega_c$, are actually welcome as they help reduce the excess early ISW power typically associated to these models). As emphasized in the recent Ref.~\cite{Poulin:2024ken} (see also Refs.~\cite{Clark:2021hlo,Vagnozzi:2023nrq}), additional late-time new physics may help in this sense by weakening the constraints on $\Omega_m$ and/or $\sigma_8$. Although realizing a successful combination of early-plus-late new physics remains non-trivial, our work and Ref.~\cite{Poulin:2024ken} (as well as the earlier attempt in Ref.~\cite{Toda:2024ncp}) emphasize the key role played by $\Omega_m$ in building such a combination. It is worth noting that, if the recent DESI BAO data are taken at face value, this task becomes somewhat easier.

Our work has also underscored the crucial importance of a consistent $\Omega_m$ calibration. At present, different late-time probes of $\Omega_m$ are in mild disagreement between each other, with recent SNeIa catalogs pushing towards increasingly high values of $\Omega_m$: focusing only on the \textit{Pantheon} versus \textit{PantheonPlus} samples, we have explicitly demonstrated the effects of a different $\Omega_m$ calibration in exacerbating the aforementioned shifts in $\omega_c$ and $S_8$. Moving forward, we therefore believe that obtaining a reliable calibration of $\Omega_m$ which is as model-independent as possible should be a key priority in the cosmology community (see also Ref.~\cite{Baryakhtar:2024rky}), in order to nail down one of the key actors in the Hubble tension play. We have argued that gas mass fraction measurements from galaxy clusters could be a promising probe in this sense, and we encourage the broader cosmology community to explore these and other less widely used probes. As we have officially entered the era of Stage IV cosmology and a solution to the Hubble tension continues to elude us, we note that the shifts in cosmological parameters which we have discussed (and which we have argued unavoidably accompany successful resolutions) inevitably leave their signatures in other cosmological observables, which current and upcoming cosmological surveys will hopefully be poised to detect~\cite{Amendola:2016saw,CMB-S4:2016ple,DESI:2016fyo,SimonsObservatory:2018koc,SimonsObservatory:2019qwx}.

\section*{Note added}

While this work was nearing completion, we became aware of the related Ref.~\cite{Poulin:2024ken} which explores similar aspects to our work, also pointing out the importance of $\Omega_m$ and $\omega_c$. We note that, despite the differences in methodology, our results are in good agreement.

\begin{acknowledgments}
\noindent S.V.\ acknowledges helpful discussions with Vivian Poulin, which originally inspired this work, and with Vittorio Ghirardini on gas mass fraction measurements. D.P., S.S.C., and S.V.\ acknowledge support from the Istituto Nazionale di Fisica Nucleare (INFN) through the Commissione Scientifica Nazionale 4 (CSN4) Iniziativa Specifica ``Quantum Fields in Gravity, Cosmology and Black Holes'' (FLAG). J.-Q.J.\ acknowledges support from the Joint PhD Training program of the University of Chinese Academy of Sciences. S.S.C.\ acknowledges support from the Fondazione Cassa di Risparmio di Trento e Rovereto (CARITRO Foundation) through a Caritro Fellowship (project ``Inflation and dark sector physics in light of next-generation cosmological surveys''). S.V.\ acknowledges support from the University of Trento and the Provincia Autonoma di Trento (PAT, Autonomous Province of Trento) through the UniTrento Internal Call for Research 2023 grant ``Searching for Dark Energy off the beaten track'' (DARKTRACK, grant agreement no.\ E63C22000500003). This publication is based upon work from the COST Action CA21136 ``Addressing observational tensions in cosmology with systematics and fundamental physics'' (CosmoVerse), supported by COST (European Cooperation in Science and Technology).
\end{acknowledgments}

\bibliography{omegac}

\end{document}